\documentclass[twocolumn, trackchanges]{aas_sample/aastex631}

\usepackage{amsmath}
\usepackage{graphicx}
\usepackage{sidecap}
\sidecaptionvpos{figure}{t}
\usepackage{longtable}
\usepackage{tabularx}
\usepackage{soul}

\begin{document}
\turnoffeditone

\title{The Coupled Impacts of Atmospheric Composition and Obliquity on the Climate Dynamics of TRAPPIST-1e}
\author{Tobi Hammond}
\affiliation{Department of Astronomy, University of Maryland, 4296 Stadium Drive, College Park, MD 20742, USA}
\author{Thaddeus Komacek}
\affiliation{Department of Astronomy, University of Maryland, 4296 Stadium Drive, College Park, MD 20742, USA}

\begin{abstract}
Planets in multi-planet systems are expected to migrate inward as near-resonant chains, thus allowing them to undergo gravitational planet-planet interactions and possibly maintain a non-zero obliquity. The TRAPPIST-1 system is in such a near-resonant configuration, making it plausible that TRAPPIST-1e has a non-zero obliquity. In this work, we use the ExoCAM GCM to study the possible climates of TRAPPIST-1e at varying obliquities and atmospheric compositions. We vary obliquity from 0$^\circ$ to 90$^\circ$ and the partial pressure of carbon dioxide from 0.0004 bars (modern Earth-like) to 1 bar. We find that models with a higher obliquity are hotter overall and have a smaller day-night temperature contrast than the lower obliquity models, which is consistent with previous studies. Most significantly, the super-rotating high-altitude jet becomes sub-rotating at high obliquity, thus impacting cloud and surface temperature patterns. As the amount of carbon dioxide increases, the climate of TRAPPIST-1e becomes hotter, cloudier, and less variable. From modeled thermal phase curves, we find that the impact of obliquity could potentially have observable consequences due to the effect of cloud coverage on the outgoing longwave radiation.
\end{abstract}

\section{Introduction} \label{Intro}
With the launch of JWST, it is now possible to characterize the atmospheres of rocky exoplanets. Planets orbiting M-dwarf stars are prime observational candidates since M-dwarfs make up about 75\% of the stellar population \citep{Henry_2004}. M-dwarf planets are easier to detect through transit and radial velocity methods and characterize by transmission spectroscopy, secondary eclipse spectroscopy, and phase curves because their host stars are less massive \citep{Nutzman_2008}. These planets are expected to be in a state of 1:1 spin synchronization due to their smaller semi-major axes at a given equilibrium temperature \citep{Pierrehumbert_2010}. This type of rotational state challenges our Earth-centric view of habitability and, more specifically, necessitates revisions to the limits of the habitable zone \citep{Kopparapu_2017}.

Clouds are critical for planetary habitability because they reflect incoming stellar irradiation to space on the dayside and inhibit radiative cooling on the nightside, especially on water-rich planets \citep{Kitzmann_2010}. \citet{Yang_2013} suggests that water clouds act as a negative feedback with increasing instellation, cooling the surface of synchronously rotating planets close to the inner edge of the HZ. Such a dayside cloud deck could potentially also be inferred from thermal phase curves since clouds absorb and re-emit thermal radiation at cooler cloud top temperatures, causing the cloudy dayside to have lower outgoing longwave radiation than cloud-free regions.

The rotation rate of tidally locked planets is a key factor for atmospheric circulation and the climate state. Following Kepler's 3rd law, the rotation rate scales with the planet's distance from the host star as $\Omega \propto a^{-3/2}$. Since cooler stars are more likely to host closer in planets, the rotation rate of synchronously rotating planets then increases with decreasing stellar type and stellar temperature. \cite{Haqq-Misra_2018} defines three regimes of planetary rotation for tidally locked planets: rapid, Rhines, and slow rotators. The Coriolis effect is weak on slowly rotating planets, resulting in a strong day-to-night atmospheric flow and a weak or nonexistent equatorial jet. Quasi-stationary clouds present at the substellar point increase the planetary albedo and cool the surface. Conversely, rapidly rotating planets have higher horizontal temperature contrasts, and their atmospheres are characterized by a strong eastward superrotating equatorial jet. This jet transports clouds eastward from the substellar point to the limb. The Rhines rotator regime, which occurs at intermediate rotation between fast and slow rotators, is characterized by planets exhibiting both strong upper-atmosphere jets and day-to-night flow with upwelling at the substellar point. This characterization of rotation regimes is also influenced by atmospheric mass, as atmospheric thickness can alter the day-night temperature contrast \citep{Chemke_2017}.

Seasonality, caused by the Earth's axial tilt, or obliquity, is one of the important factors that facilitate habitability \citep{Williams_1997, Barnett_2022}. A non-zero planetary obliquity allows for the location of the point of maximum stellar irradiation to change over the course of one orbit, which in turn increases the amount of habitable area on the planet \citep{Dobrovolskis_2009}. Planets with an orbital period equal to their rotation period and a non-zero obliquity will be forced out of their tidally locked state and experience seasonal cycles. Previous studies \citep{Dobrovolskis_2009, Wang_2016, Rauscher_2017, Chen_2023} for non-synchronously rotating planets around Sun-like stars indicate that increasing the obliquity of a planet can increase the outer edge of the habitable zone. For planets with an obliquity above 54$^{\circ}$, polar temperatures are higher, and thus it is more difficult for the planet to be globally frozen \citep{Williams_2003, Armstrong_2014, Kilic_2017}. As a result, it is possible that high-obliquity planets may be more habitable due to more globally uniform surface temperatures \citep{Olson_2018, Jernigan_2023}. 

Previous work has suggested that the same tidal and gravitational forces that spin-synchronize a planet should also damp the planet's obliquity to zero \citep{Heller_2011}. However, \cite{Millholland_2019} shows that an outer companion can gravitationally perturb a planet, allowing it to maintain a non-zero obliquity. Capture in a so-called ``Cassini State'' allows planets in multi-planet systems to persistently gravitationally perturb each other so that they are locked in near-resonant configurations \citep{Millholland_2019, Su_2022}. Specifically, the second Cassini state drives orbital precession through tidal forces. This acts to reorient the planet's rotational axis and maintain non-zero obliquities over long timescales \citep{Peale_1969}. Multi-planet resonant chains also facilitate inward migration \citep{Vinson_2017,Millholland_2019}, which is favorable for maintaining surface liquid water.

Previous studies \citep{Showman_2013, Kopparapu_2017, komacek_2019, Yang_2023} often assume a 0$^\circ$ obliquity state when investigating the atmospheric dynamics of synchronously rotating terrestrial planets around M-dwarf stars. However, some studies have relaxed the assumption of $0^\circ$ obliquity. \cite{Wang_2016} uses a 3D GCM to model the effects of obliquity on habitability for GJ 667Cc at varying instellation. They find that the width of the habitable zone decreases as obliquity increases for a slowly rotating planet. \cite{Ohno_2019, Ohno_2019_2} define five dynamical regimes for models with non-zero obliquities and eccentricities. These regimes are determined by comparing the orbital ($P_{orb}$) and rotational ($P_{rot}$) periods to the radiative timescale ($\tau_{rad}$). \edit1{The radiative timescale 
generally indicates the strength of seasonal cycles and efficiency of atmospheric circulation \citep{Guendelman_2019}. \cite{Ohno_2019} use this parameter to characterize how the stellar irradiation and rotation rate affect the resulting seasonal variations of temperature. For example, when the stellar irradiation is very strong and the atmospheric response time is shorter than the rotation period (Regime I, $\tau_{rad} < P_{rot}$), the temperature patterns show a strong day-night contrast. Alternatively, when the stellar irradiation is very weak and the atmospheric response time is much longer than the orbital period, the temperature patterns are instead dependent on the annual mean forcing (Regimes IV and V, $\tau_{rad} > P_{orb}$). In between these regimes (i.e., $P_{rot} < \tau_{rad} < P_{orb}$, Regimes II and III), the temperature patterns are determined by the diurnal mean forcing. For the purposes of this work, the radiative timescale is estimated as \citep{Seager_2010}
\begin{equation}
\label{eq:trad}
\tau_{rad} \approx\frac{p}{g}\frac{c_p}{4\sigma T_s^3},
\end{equation}
where $p$ is the surface pressure, $\sigma$ is the Stefan-Boltzmann constant, $c_p$ is the specific heat for an assumed atmospheric composition \citep{Pierrehumbert_2010}, $T_s$ is the globally averaged surface temperature, and $g$ = 8.01 ms$^{-2}$.} Since \edit1{temperate rocky} planets orbiting M-dwarf stars have very weak stellar irradiation, their temperature and wind patterns are \edit1{likely} dominated by the annual mean instellation \edit1{per \cite{Ohno_2019}}. This climate state corresponds to their regimes IV and V ($\tau_{rad} > P_{orb}$). Regime IV corresponds to obliquities less than 54$^\circ$ \citep{Ward_1974} and is characterized by a hot equatorial region and eastward winds. Conversely, Regime V corresponds to obliquities greater than 54$^\circ$ and is characterized by hot polar regions and westward winds. \cite{Ohno_2019} use an idealized shallow water model, so we will test their results using a robust 3D climate model. 

In this study, we use the 3D ExoCAM GCM to explore the implications of varying obliquity and atmospheric composition. We chose to model one of the best rocky planet candidates for atmospheric characterization, TRAPPIST-1e \citep{Gillon_2017, Agol_2021}. TRAPPIST-1 is a cool M-dwarf star hosting seven known planets \citep{Gillon_2017}. This system is in a near-resonant configuration \citep{Luger_2017}, making it likely that its planets have non-zero obliquities \citep{Guerrero_2023}. Additionally, TRAPPIST-1e is the most well-studied planet in the system by previous 3D GCMs \citep[e.g.,][]{Sergeev_2022}, as it is the only planet in the system that is robustly habitable assuming Earth-like conditions \citep{Wolf_2017}.

This paper is organized as follows. In Section \ref{Models} we describe the model setup and present our parameter choices and data analysis methods. In Section \ref{Results} we show our results for simulations of both varying obliquity and partial pressure of CO$_2$. In Section \ref{Discussion}, we discuss the implication of this analysis and how it compares to previous work.  Finally, we present our conclusions and summary in Section \ref{Conclusion}.
\section{Models} \label{Models}
\subsection{ExoCAM Setup}
We used ExoCAM\footnote{https://github.com/storyofthewolf/ExoCAM}, a 3D exoplanet general circulation model (GCM) \citep{Wolf_2022} to explore the climate of TRAPPIST-1e. ExoCAM is a modified version of the Community Earth System Model (CESM) version 1.2.1 \citep{neale_2012} that adds ExoRT\footnote{https://github.com/storyofthewolf/ExoRT}, a non-gray correlated-k radiative transfer scheme. ExoCAM is a widely established and accessible model used to study the climate and atmospheric properties of exoplanets \citep{Kopparapu_2017, Haqq-Misra_2018, komacek_2019, May_2021, Lobo_2023}, including the TRAPPIST-1 system \citep{Wolf_2017, Hu_2020, Rushby_2020, THAI_2, Hochman_2022, Rotman_2023}. 

We assume in our models that the surface is an aqua planet covered entirely by a slab ocean with a depth of 50 m while ignoring ocean heat transport (i.e., zero Q-flux). We allow sea ice to form, with its thickness and distribution dictated by the CESM thermodynamic sea ice scheme \citep{Bitz_2012}. The water vapor mixing ratio, or humidity, is also spatially and temporally variable. The Clausius-Clapeyron relation determines its near-surface abundance, with circulation and condensation also determining its abundance in the rest of the atmosphere. We include both liquid and ice water clouds, assuming liquid clouds have a fixed particle radius \citep{komacek_2019} and ice cloud particles have a temperature-dependent mean radius. 

\subsection{Model Grid} 
For all models, we use the updated planetary parameters of TRAPPIST-1e from \cite{Agol_2021}. We assume the radius of TRAPPIST-1e to be 0.92 $R_\oplus$, the surface gravity to be 8.01 ms$^{-2}$, and the incident stellar flux as 879 Wm$^{-2}$. We use the corresponding incident stellar spectra for a 2600 K star from \cite{Allard_2007}. Additionally, all cases maintain the modern Earth-like parameters of 1 bar N$_2$ and $1.7 \times ~10^{-6}$ bar CH$_4$. For simplicity, we assume that TRAPPIST-1e is in 1:1 spin-synchronization with its host star, thus setting its orbital and rotational periods equal to 6.0997 days. The eccentricity is also set to zero for all cases.

We conduct a grid of varying partial pressures of carbon dioxide and obliquities for TRAPPIST-1e. Specifically, we conduct simulations for partial pressures of 1 bar, 0.1 bar, 0.01 bar, and 0.0004 (Earth-like) bar CO$_2$; each with obliquities of 0$^\circ$, 22.5$^\circ$, 45$^\circ$, 67.5$^\circ$, and 90$^\circ$ (Table \ref{tab:table}). Note that at 54$^\circ$, the planet begins to receive more instellation at the pole than at the equator \citep{Ohno_2019} which has significant effects on climate, as we will show. Following \citet{Wolf_2017}, the range of pCO$_2$ levels was chosen to cover possible habitable climate states of TRAPPIST-1e. Here, the lowest value was chosen to resemble modern Earth's pCO$_2$ level with a relatively cold climate state, while the highest case is to resemble a hot-house climate state. 

\begin{table}[!ht]
    \centering
    \caption{Model parameters varied in this study. }
    \label{tab:table}
    \begin{tabular}{|l|r|} \hline  
         \bf{Parameter}&  \bf{Values}\\ \hline  
         Obliquity [$^\circ$]&  0, 22.5, 45, 67.5, 45, 90\\ \hline  
         pCO$_2$ [bar]&  0.0004, 0.01, 0.1, 1\\ \hline
         pN$_2$ [bar]& 1\\ \hline
         pCH$_4$ [bar]& $1.7 \times ~10^{-6}$\\ \hline
         Radius [R$_{\oplus}$]&  0.92\\ \hline
         Surface Gravity [ms$^{-2}$]&  8.01\\ \hline
         Period [Earth Days]& 6.0997\\ \hline
         Eccentricity & 0\\ \hline
         Stellar Flux [Wm$^{-2}$]& 879\\ \hline
         Stellar Temperature [K]& 2600\\\hline
    \end{tabular}
\end{table}

\subsection{Numerical Details} 
We use a 4$^\circ$ x 5$^\circ$ horizontal resolution grid with a finite volume dynamical core and 40 atmospheric levels from the surface to a pressure of 1 mbar. Each simulation uses a 30-minute dynamical time step and a 60-minute radiative time step and was run until equilibrium. This was determined to be when both the top-of-atmosphere net radiation balance and the mean surface temperature reached a steady state, which was generally after 40–50 Earth years. \edit1{Given the 6.1 day period of TRAPPIST-1e, we predict that all models will have radiative timescales longer than their orbital periods and are thus in the regime where annual mean forcing is dominant \citep{Ohno_2019}.} Therefore, all mean quantities shown here were taken using the time average of the last 10 Earth years of model output data. 

\section{Results} \label{Results}
\subsection{Mean Climate}  
\begin{figure*}[ht]
    \centering
    \includegraphics[width=0.8\textwidth]{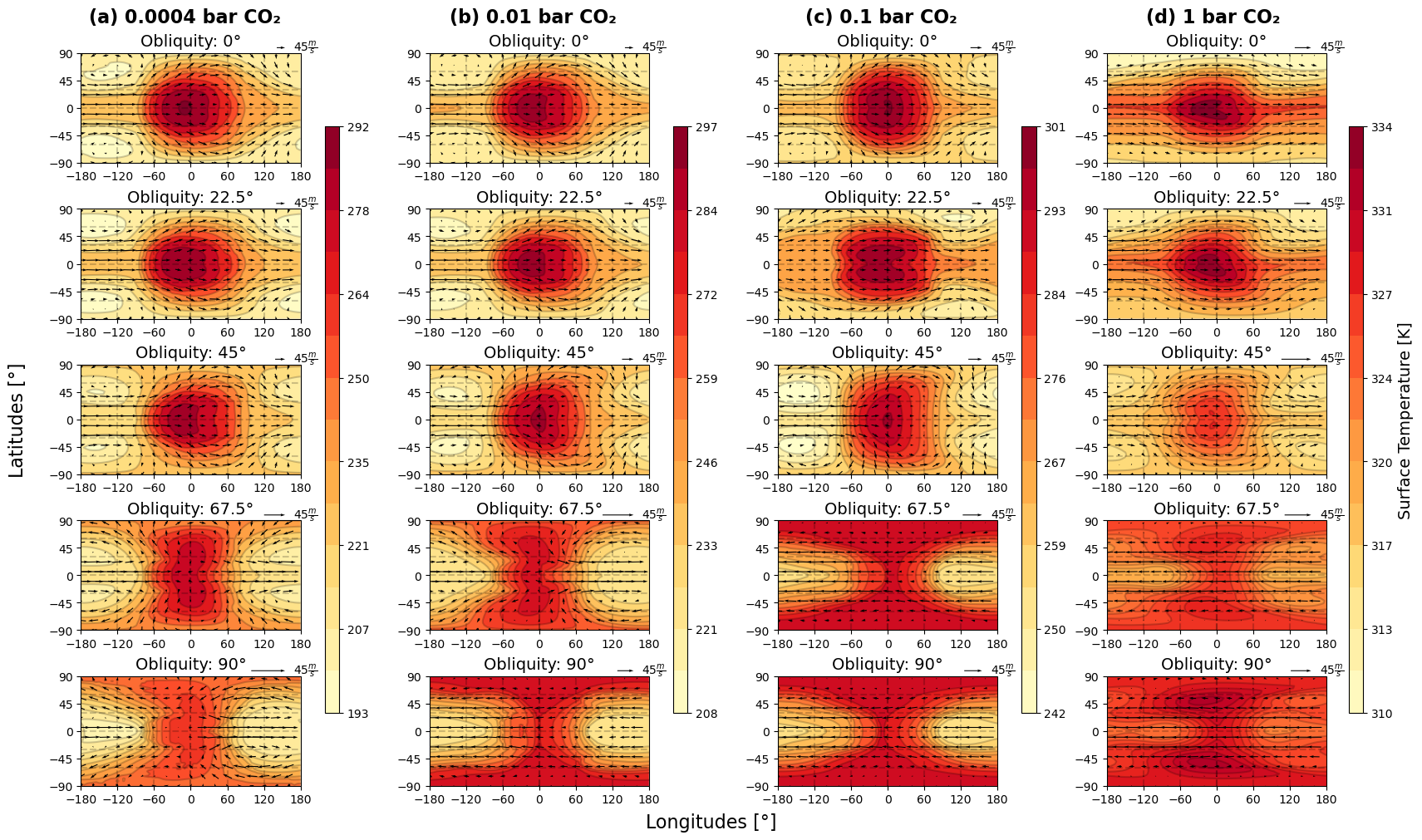}
    \caption{Maps of surface temperature (colors) and 430 hPa winds (arrows) for each GCM simulation with all CO$_2$ cases (each column) having independent color bars. The substellar point is located at (0$^\circ$, 0$^\circ$). The substellar point is hottest at lower obliquities, and the poles are hotter than the equator at higher obliquities. Day-night surface temperature contrasts and resulting planetary-scale atmospheric wave patterns, like Rossby gyres, are more prominent for the cases with less CO$_2$. For the higher obliquity cases the direction of the equatorial wind speed reverses from eastward to westward.}
    \label{fig:surface_temps}
\end{figure*}

\begin{figure*}[ht]
    \centering
    \includegraphics[width=0.8\textwidth]{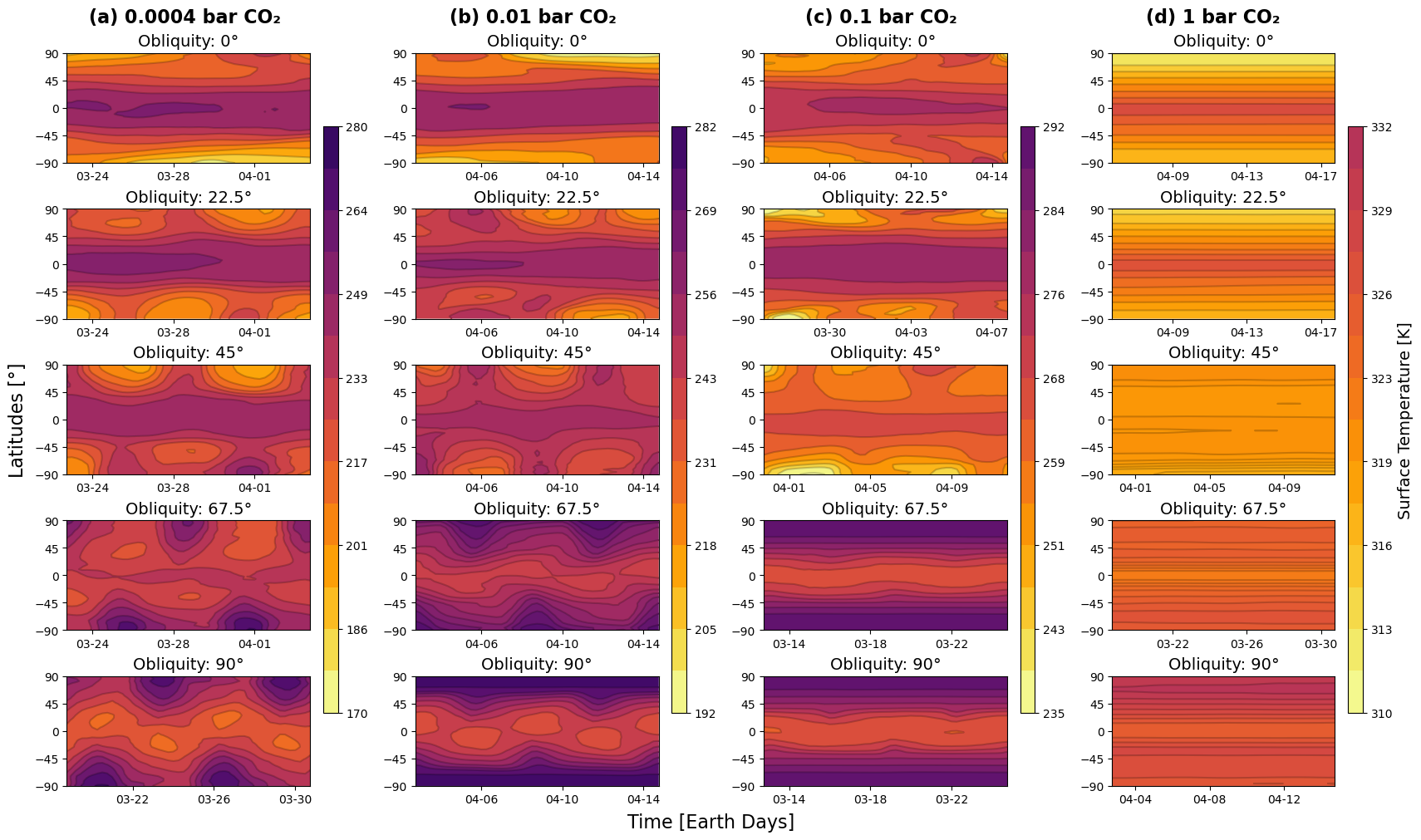}
    \caption{Maps of surface temperature (colors) over the last \textbf{~12 days} of model time for each GCM simulation. The time axis shows the actual model time in Earth days. All CO$_2$ cases (each column) have independent color bars. Cases with hotter surface temperatures dampen seasonal variability in favor of long-term atmospheric variation.}
    \label{fig:time_st}
\end{figure*}

To display the overall effect of obliquity on the climate of TRAPPIST-1e, we show maps of the time-mean surface temperature for each model case in Figure \ref{fig:surface_temps}. 
Since only one hemisphere is illuminated by the host star, the surface temperature patterns for the zero obliquity cases resemble an ``eyeball planet,'' \citep{Pierrehumbert_2010} with high day-night and equator-to-pole temperature contrasts. For cases with a non-zero obliquity, the substellar point migrates north and south over the course of one orbital period, with the latitude of maximum deviation from the equator being the planet's obliquity \citep{Dobrovolskis_2009}. To track the atmospheric response to this variation in irradiation, we plot the surface temperature over time in Figure \ref{fig:time_st}.
Generally, the path of maximum instellation is translated in the surface temperature, but the overall seasonal variation is dominated by the ratio between the radiative timescale and the orbital period \citep{Ohno_2019}. 
\begin{figure}[htpb]
    \includegraphics[width=0.47\textwidth]{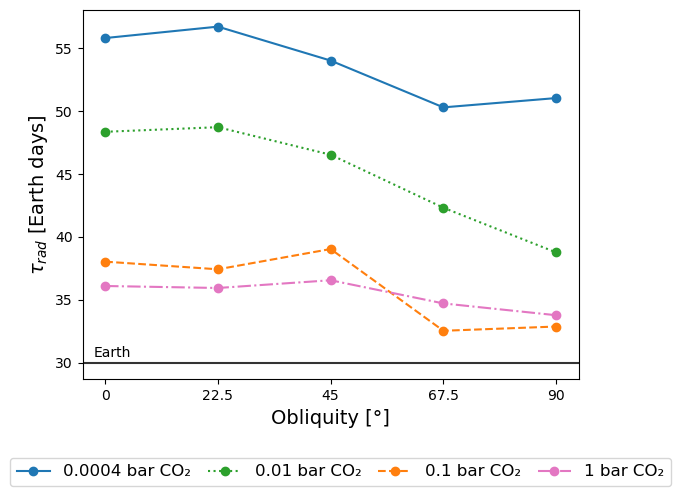}
    \centering
    \caption{Plots of the radiative timescale ($\tau_{rad}$) against obliquity for each case. Each of $c_p$, $p$, and $T_s$ are varied accordingly for each pCO$_2$ case. The solid black line indicates Earth's radiative timescale ($\sim$30 days, \citealp{Guendelman_2022}). As expected, we find that for all cases $\tau_{rad}$ is larger than the orbital period. Additionally, $\tau_{rad}$ generally increases with decreasing surface temperature.} 
    \label{fig:rad_timescale}
\end{figure}

\edit1{To determine the approximate dynamical regime of each planet, \citep{Ohno_2019}, we estimate the radiative timescale using Equation (\ref{eq:trad}), varying $c_p$, $p$, and $T_s$ for each pCO$_2$ case, and show the results in Figure  \ref{fig:rad_timescale}. As predicted, the radiative timescale for all cases is much longer than TRAPPIST-1e's orbital period (6.1 days), indicating that all cases should be in the annual mean forcing regime. Since the radiative timescale scales with $1/T_s^3$, we find that the hottest simulations with higher pCO$_2$ values generally have the shortest radiative timescales.} \edit1{The atmospheric composition and host star type are also expected to impact seasonality, which are neglected in our simple estimation of the radiative timescale \citep{Tan_2022}}. For example, the map of the 1 bar pCO$_2$ case with \edit1{45$^\circ$} obliquity appears asymmetrical in this short snapshot -- a distinguishable pattern only appears after multiple orbital/rotational periods.

We also see in Figures \ref{fig:surface_temps} and \ref{fig:time_st} that as obliquity increases, the equator-to-pole contrast decreases. Between 45$^\circ$ and 67.5$^\circ$ obliquity, there is a distinct inflection point where the poles become hotter than the equator, consistent with the threshold obliquity of 54$^\circ$ \citep{Ward_1974}. These variations in mean climate with obliquity are most noticeable as pCO$_2$ increases, due to increased atmospheric moisture warming the deep atmosphere via both the 
greenhouse effect and enhanced absorption of incident starlight by water vapor.

\begin{figure}[htpb]
    \includegraphics[width=0.47\textwidth]{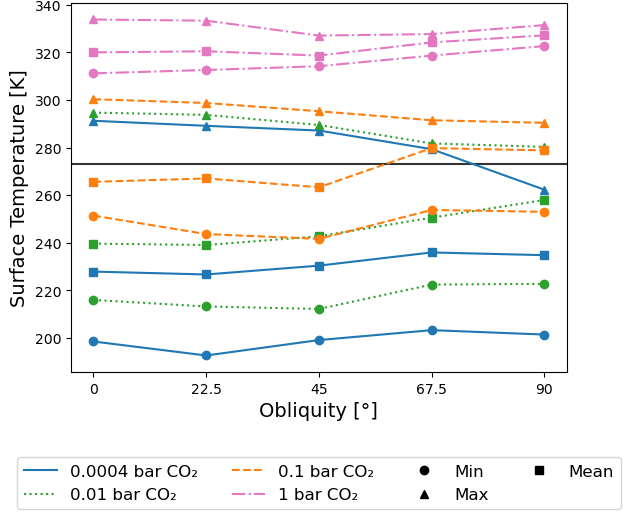}
    \centering
    \caption{Plots of \textbf{time averaged global} maximum, minimum, and mean surface temperature against obliquity for each case The solid black line indicates the melting point of water. As expected, we find that lower pCO$_2$ causes a lower mean surface temperature with obliquity having a secondary effect. Overall, higher obliquity cases have lower temperature differences between cases with high and low pCO$_2$.} \label{fig:min_max_mean_st}
\end{figure}

In Figure \ref{fig:min_max_mean_st}, we plot the \edit1{time averaged global} minimum, mean, and maximum surface temperatures for each pCO$_2$ case with respect to obliquity. \edit1{Since all cases are synchronously rotating, generally the minimum value corresponds to the mean minimum night side temperature, and the maximum value corresponds to the mean peak dayside temperature, especially for low obliquity cases. Note that the ``Earth-like'' case ($4 \times 10^{-4}$ bar CO$_2$, 22.5$^\circ$ obliquity) is not expected to have Earth-like temperatures due to the difference in instellation and the the assumption of synchronous rotation.}


Similar to \cite{Wolf_2017}, the climate state and surface temperature are dependent on the partial pressure of CO$_2$. Here we can see that the average surface temperature increases and the surface temperature range decreases with increasing pCO$_2$ due to the increasing greenhouse effect. From the distribution of the mean surface temperatures, it is clear that the changes due to varying pCO$_2$ for mean surface temperature are greater than the effects due to obliquity. There is also not a consistent trend from obliquity across the various pCO$_2$ regimes, although, the surface temperature range generally decreases with increasing obliquity. This is consistent with our finding that increasing obliquity decreases the equator-pole temperature contrast. 
\subsection{Circulation}
\begin{figure*}[ht]
    \centering
    \includegraphics[width=0.75\linewidth]{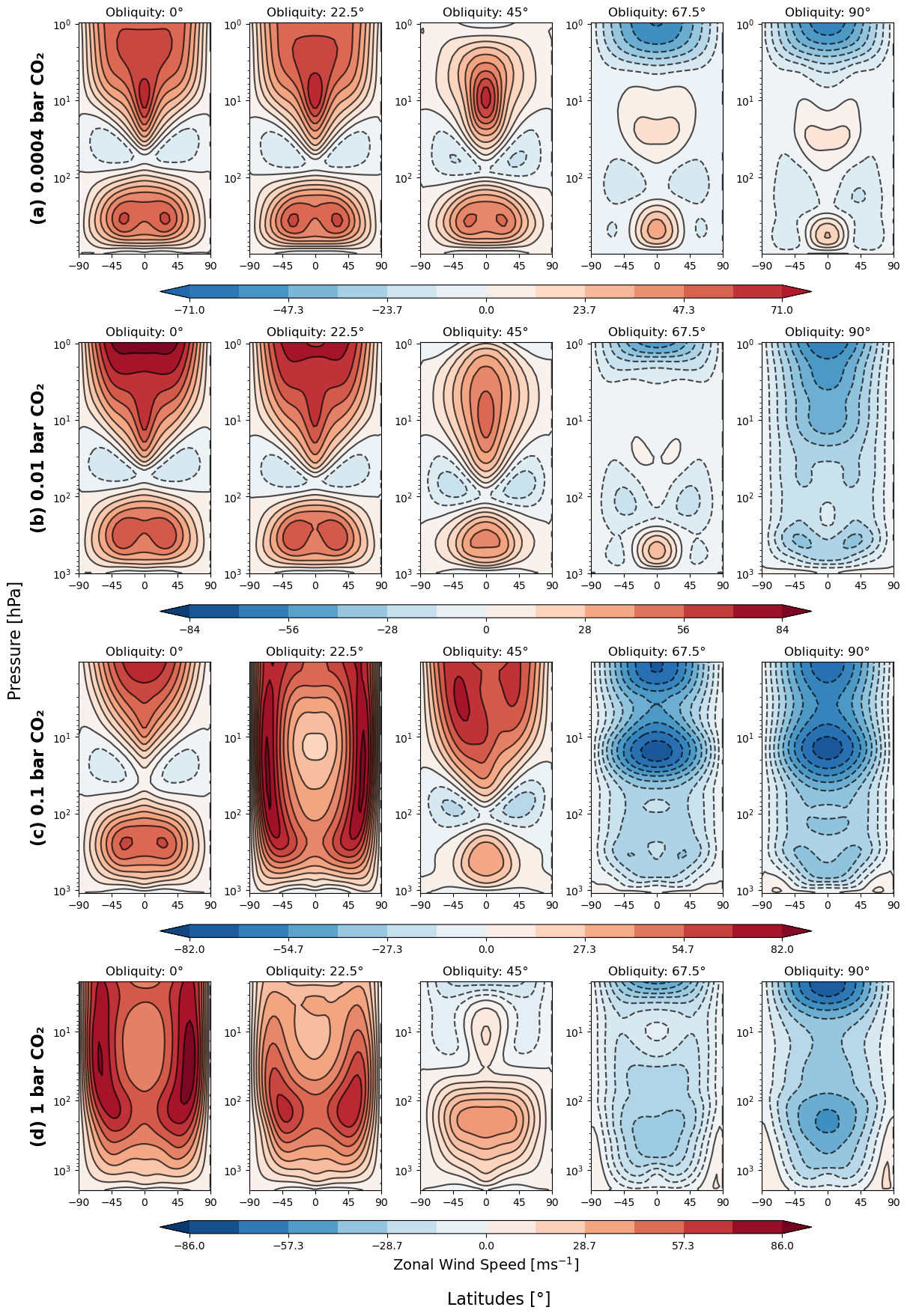}
    \caption{Maps of 
    zonal mean zonal wind speeds (colors and contours). Negative (blue) values indicate 
    westward motion, while positive (red) values indicate 
    eastward motion. For the contours of zonal wind, solid lines indicate eastward moving winds, while dashed lines indicate westward moving winds. The reversal of equatorial wind speeds when the obliquity changes from 45$^\circ$ to 67.5$^\circ$ is seen here as well.}
    \label{fig:zonal_wind}
\end{figure*}

To supplement the near-surface horizontal wind isobars in Figure \ref{fig:surface_temps} and study the vertical structure of winds, we plot maps of the zonal-mean zonal wind structure in Figure \ref{fig:zonal_wind}.
Studying the near-surface horizontal wind, we find that for all pCO$_2$ cases at low obliquities, there is prominent eastward flow. Specifically for cooler cases, there are Rossby gyres in the mid-latitudes east of the substellar point. The eastward jet is due to heating at the substellar point, driving a planetary-scale wave pattern that fluxes momentum toward the equator via wave-mean flow interactions \citep{Showman_2013}. We also see eastward flow in the plots of zonal-mean zonal wind that gets stronger as pCO$_2$ increases.

At the 54$^\circ$ boundary, the wind direction flips to dominantly westward flow in the free atmosphere, as predicted by \cite{Ohno_2019}. In Figure \ref{fig:surface_temps}, the wind arrows switch from pointing right to pointing left, and in Figure \ref{fig:zonal_wind}, the zonal wind contours change from red (solid lines) to blue (dashed lines). Above 54$^\circ$ for all CO$_2$ cases, the upper atmospheric jets become stronger and wind speeds generally increase.

\begin{figure*}[ht!]
    \centering
    \includegraphics[width=\textwidth]{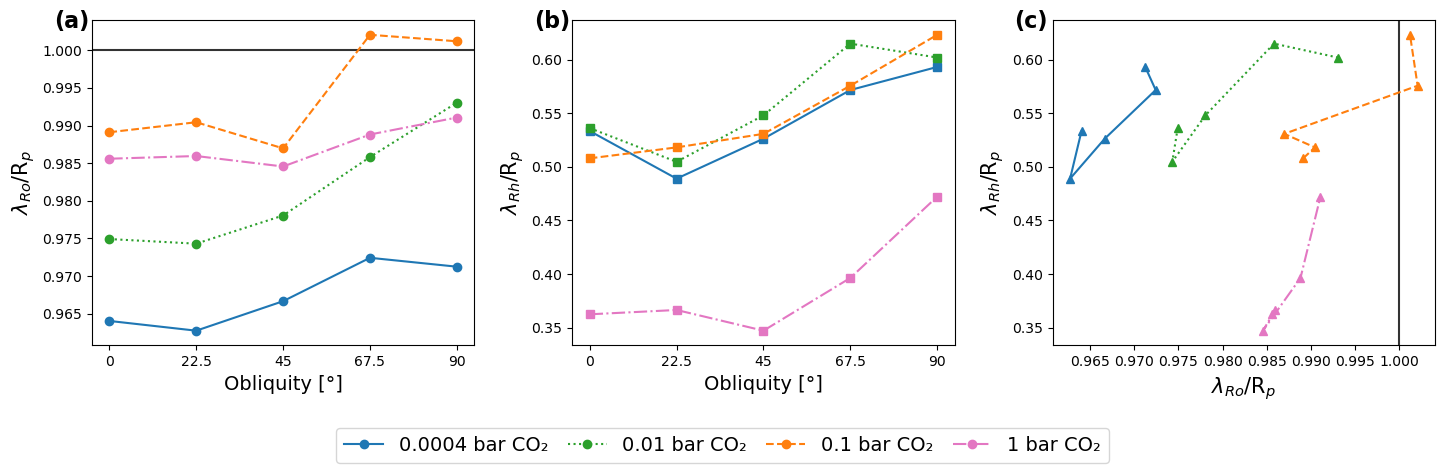}
    \caption{Plots of non-dimensional Rossby deformation radius ($\lambda_{Ro}$/$R_p$) versus obliquity (a), non-dimensional Rhines length ($\lambda_{Rh}$/$R_p$) versus obliquity (b), and non-dimensional Rhines length versus non-dimensional Rossby deformation radius (c) for each pCO$_2$ case. Rapid rotators are defined when $\lambda_{Rh}$/$R_p$ $<$ 1 and the $\lambda_{Ro}$/$R_p$ $<$ 1. Rhines rotators are defined when $\lambda_{Rh}$/$R_p$ $<$ 1 and $\lambda_{Ro}$/$R_p$ $>$ 1. Here most cases are within the rapid rotator regime except for the high obliquity 0.1 bar CO$_2$ cases, which are in the Rhines rotator regimes.}
    \label{fig:rossby_def}
\end{figure*}

\cite{Haqq-Misra_2018} calculate that TRAPPIST-1e is in between the rapid rotator and Rhines rotator regimes, and thus its global circulation could exhibit aspects of both. To confirm their findings for cases with varying pCO$_2$ and obliquity, we calculate the equatorial Rossby deformation radius and Rhines length for each case in Figure \ref{fig:rossby_def}. We calculated the equatorial Rossby deformation radius following \cite{Haqq-Misra_2018} as:
\begin{equation}
    \label{eq:Rossby}
    \lambda_{Ro} = \sqrt{\frac{R_p\sqrt{gH}}{4\Omega}}~\mathrm{.}
\end{equation}
In Equation (\ref{eq:Rossby}), $\Omega$ is TRAPPIST-1e's rotation rate at the equator, H is the atmospheric scale height, $R_p$ = 0.92 $R_\oplus$, and $g$ = 8.01 ms$^{-2}$. We set H = $T_sR/m_{air}g$, where $T_s$ is the globally averaged surface temperature from Figure \ref{fig:min_max_mean_st}, $m_{air}$ is the mean molecular weight for each pCO$_2$ case, and R is the universal gas constant. Further, we express the Rhines length as: 
\begin{equation}
    \label{eq:Rhines}
    \lambda_{Rh} = \pi\sqrt{\frac{R_pU}{2\Omega}}~\mathrm{.}
\end{equation}
In Equation (\ref{eq:Rhines}), U is the characteristic root mean squared velocity, calculated here as the root mean squared of the globally averaged meridional wind speed at the surface. 
Most cases are indeed within the rapid rotator regime, defined when non-dimensional Rossby deformation radius ($\lambda_R$/$R_p$) is less than one \citep{Haqq-Misra_2018}. There is a general trend that $\lambda_{Ro}$/$R_p$ and $\lambda_{Rh}$/$R_p$ increase as obliquity increases. 

\subsection{Cloud Cover}
\begin{figure*}[ht]
    \centering
    \includegraphics[width=0.8\textwidth]{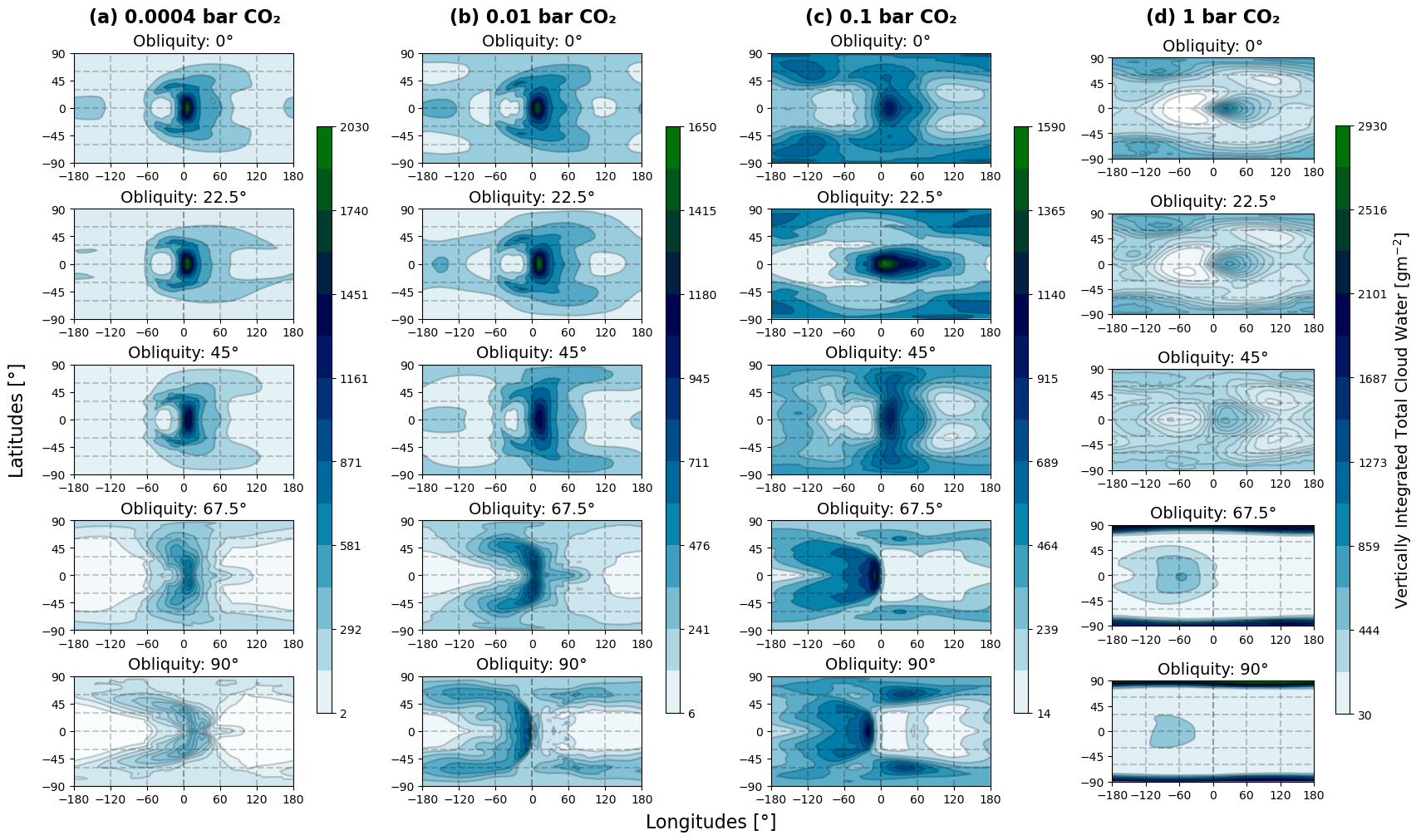}
    \caption{Maps of vertically integrated total cloud water path (colors) for each GCM simulation with all pCO$_2$  cases (each column) having independent color bars. Darker colors indicate areas with more clouds. For low obliquity cases, the cloudiest areas occur just east of the substellar point due to advection by the superrotating jets. As with surface temperature, the highest column mass of clouds are at the poles for models with higher obliquities. The change in direction of the high-altitude superrotating jet in the high obliquity cases causes the peak in cloud water path to be west of the substellar point.}
    \label{fig:cloudwater}
\end{figure*}

To study the global cloud coverage, we plot maps of the vertically integrated cloud water path for each model case in Figure \ref{fig:cloudwater}. 
As expected, the average time-mean cloud coverage increases with increasing pCO$_2$. At higher surface temperatures, more water evaporates from the ocean surface, thus increasing the mixing ratio of atmospheric water vapor that can be lofted to low pressures and then condense to form clouds.

At 0$^\circ$ degrees obliquity, the maximum time-mean cloud coverage is localized just east of the substellar point, as our simulations of TRAPPIST-1e are in the near-rapid rotator regime with a superrotating equatorial jet \citep{Wolf_2017, May_2021, Sergeev_2022}. This pattern is expected for a synchronously rotating temperate terrestrial planet \citep{Yang_2013}. As obliquity increases, the time-mean equator-to-pole cloud cover contrast decreases for low to moderate pCO$_2$ and the area of maximum cloud coverage generally remains east of the substellar point. Above 54$^\circ$, the maximum cloud coverage is instead localized west of the substellar point due to the reversal of high-altitude winds. 
\subsection{Observables}
\begin{figure*}[ht]
    \centering
    \includegraphics[width=\textwidth]{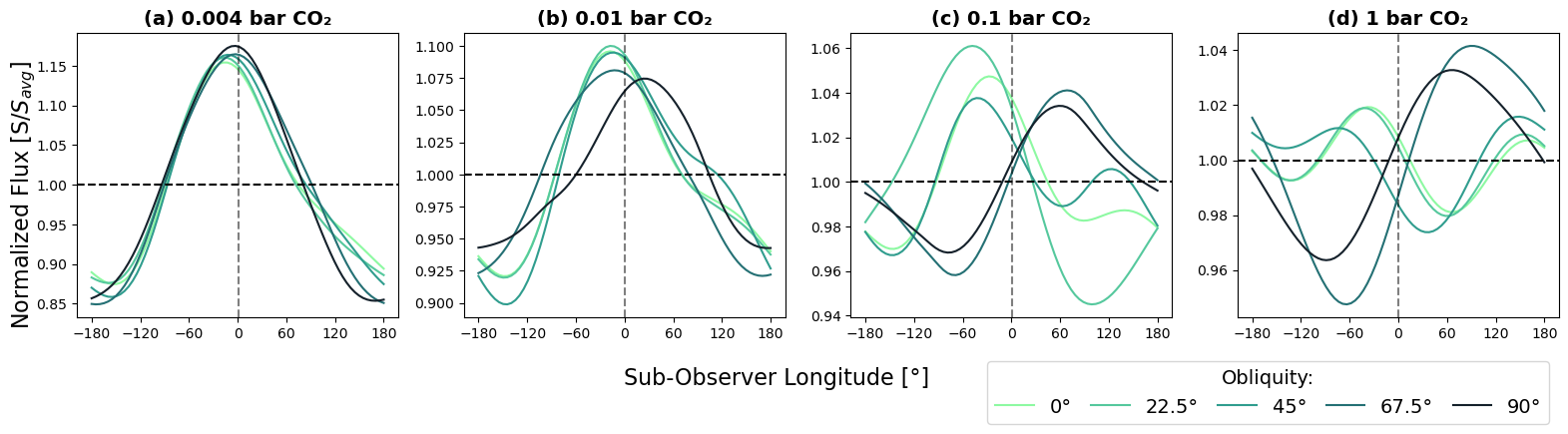}
    \caption{Simulated phase curves of the normalized outgoing longwave flux for each model case with varying pCO$_2$ (increasing from left to right) and obliquity (colors).  A sub-observer longitude of 0 corresponds to secondary eclipse 
    As pCO$_2$  increases, the phase curve amplitude decreases (note the varying y-axis scale between sub-plots). For higher pCO$_2$ cases, the period of the peak flux shifts from post- to pre-eclipse when the obliquity increases from 45$^\circ$ to 67.5$^\circ$  (darker colors).}
    \label{fig:phase}
\end{figure*}

In order to understand the observational effects of obliquity on planetary thermal emission, we constructed bolometric phase curves, shown in Figure \ref{fig:phase}.
These phase curves are calculated using the methods in \citet{Cowan_2008}. Specifically, we integrated the outgoing top-of-atmosphere time-averaged longwave flux over each latitude and longitude from hemispheres centered at each sub-observer longitude in the planet's orbit. 
We calculated the sub-observer position following \cite{Rauscher_2017} as 
\begin{equation}
    \sin(\phi_{obs}) = \sin(\psi)\sin(2\pi/P_{orb})~\mathrm{,}
\end{equation}
where $\phi_{obs}$ is the sub-observer latitude, $\psi$ is the obliquity, and $P_{orb}$ = 6.0997 days is the orbital period. This assumption corresponds to an edge-on, or 90$^\circ$ orbital inclination, relative to the observer, which is consistent with TRAPPIST-1e's observed orbital inclination \citep{Agol_2021}. We calculate the phase curves from the time averaged data since there is little seasonality in each case. Thus, we assume that at secondary eclipse, the entirety of the planet's dayside is facing the observer and in Figure \ref{fig:phase} a sub-observer longitude of 0 corresponds to secondary eclipse. 

We find that the outgoing longwave radiation peaks at the sub-observer longitude where the hemispherically integrated cloud cover is at a minimum. Clouds absorb thermal radiation escaping the atmosphere and re-emit it due to cooler cloud top temperatures, so cloud coverage inversely relates to the amount of thermal flux that can escape from the warmer deep atmosphere. As pCO$_2$, and thus the surface temperature, increases, the normalized flux amplitude decreases. Hotter cases have lower day-night temperature contrasts due to moisture enhancing the effective day-night heat transport \citep{Haqq-Misra_2018}. 
\section{Discussion} \label{Discussion}
\subsection{Comparisons to Previous Work}
We now compare our results to the limited set of studies on the impact of non-zero obliquity on the climates of synchronously rotating rocky planets orbiting late-type M-dwarf stars. The dependence of the time-mean surface temperature and instellation patterns on obliquity broadly agrees with \cite{Wang_2016} and \cite{Dobrovolskis_2009}. \cite{Wang_2016} uses a 3D GCM to model the atmospheric circulation of GJ 667Cc and study the effects of obliquity on climate, assuming the planet is in a state of synchronous rotation. GJ 667Cc has a period of 28 days, and the host star temperature is set at 3700 K. Similar to our models, they set their background atmospheric parameters as 1 bar N$_2$ and 1.7 ppmv of CH$_4$. They also set their CO$_2$ partial pressure as 355ppmv, comparable to our 0.0004 bar case. The greater rotation period and host star temperature in their simulations classify GJ 667Cc as a slow rotator \citep{Noda_2017, Haqq-Misra_2018}. As opposed to our models of TRAPPIST-1e, their results show a weak equatorial jet and strong day-to-night atmospheric flow. Given the dynamical regime difference, they don't find a drastic change in the global circulation pattern with obliquity. However, \cite{Wang_2016} finds that mean surface temperature increases with increasing obliquity. Comparing this study's 0.0004 bar case to their results, the time-mean average surface temperature slightly increases with obliquity as in \cite{Wang_2016}, but decreases at 90$^\circ$ obliquity.
\subsection{Implications for Observations and Habitability} 
Our 
orbital phase curves in Figure \ref{fig:phase} show that cases with less pCO$_2$ will have a larger phase curve amplitude due to greater day-night temperature contrasts, but also have a smaller difference in peak sub-observer longitude offset between obliquity cases due to reduced cloud cover. Thus, given sufficiently sensitive measurements with a future mission that can detect infrared phase curves of temperate rocky planets (e.g., MIRECLE, LIFE), it may be possible to infer a non-zero obliquity of a planet from its phase curve. As investigated by \cite{Rauscher_2017} and \cite{Ohno_2019_2}, orbital geometry is an important parameter to consider when interpreting the phase curves of planets with a non-zero obliquity. In this study we assume the dayside faces the observer at secondary eclipse, but in reality this is not known a priori. As discussed in \cite{Rauscher_2017}, viewing angle plays a large role in the shape of the phase curve. The vast majority of nearby temperate rocky planets are non-transiting \citep{Kossakowski_2023}, therefore thermal phase curves at a range of viewing angles can provide information about obliquity. The Planetary Infrared Excess (PIE) technique can provide insight into the atmospheric characteristics of nearby, non-transiting exoplanets \citep{Stevenson_2020}. 

\cite{Wolf_2017} explores the possible habitability of planets in the TRAPPIST-1 system over a range of atmospheric compositions. They find that TRAPPIST-1e is robustly in the habitable zone and that their models of TRAPPIST-1e are habitable for pCO$_2$ values up to 1 bar with a 1 bar N$_2$ background gas composition. In agreement with \cite{Wolf_2017}, we find that the maximum time-mean surface temperature is above 273K at zero obliquity for all CO$_2$ cases (Figure \ref{fig:min_max_mean_st}). 
As obliquity increases, the maximum surface temperature stays above 273 K in most cases, which is generally more favorable for habitability. However, all cases experience relatively weak intra-seasonal variations due to the short orbital period of TRAPPIST-1e. Since seasonality is an important factor for life on Earth, future work could explore the biochemical impacts of weak seasonality of short-period planets with non-zero obliquities. 

\subsection{Limitations and Future work}
To fully characterize temperate rocky exoplanets, it is necessary to account for ocean dynamics, ocean heat transport, and sea ice drift \citep{Shields_2013}. For the sake of simplicity, we chose to neglect these variables in order to conduct a broad suite of 20 GCMs. However, ocean dynamics have been shown to affect the global surface temperature patterns and resulting brightness patterns of rocky planets. Heat transport from the dayside to the nightside leads to the formation of a ``lobster'' pattern of sea surface temperature rather than an ``eyeball'' pattern \citep{Hu_2014}.

We also did not consider the effects of continents, which will affect the water vapor abundance and cloud coverage \citep{Lewis_2018, Salazar_2020}. Continents on tidally locked planets are predicted to align with the antistellar or substellar points due to true polar wander \citep{Leconte_2013}. A non-zero obliquity could cause the continent to be located further from the substellar point, as it will align to be furthest from the rotation axis, which is no longer necessarily perpendicular to the orbital plane. The combination of continents and ocean dynamics will incite biomass transport from the surface to the ocean floor  \citep{libes_2011}. This can be an important parameter when considering planetary habitability and the possibility of photosynthetic life \citep{Yang_2019, Olson_2020}. 

In addition to continents and ocean dynamics, varying eccentricity and orbital spin rate could impact the climate and seasonality of planets with a non-zero obliquity \citep{Dobrovolskis_2013, Leconte_2015}. Planet-planet interactions in compact multi-planet systems can cause chaotic spin variations, which affect climate evolution and stability. \cite{Chen_2023} use an N-rigid-body simulation to investigate the effects of spin-orbit variations for TRAPPIST-1e and f. They find that these variations drive the planets out of spin-synchronization, but the climatic influences are less extreme for closer in planets (i.e., TRAPPIST-1e). Although \cite{Ohno_2019} suggests that seasonal effects due to eccentricity are weak for planets in regimes IV and V, where the radiative timescale is longer than an orbital period, it would be interesting to compare their results to 
less idealized models. \cite{Jernigan_2023} shows that for non-synchronously rotating planets around Sun-like stars, the coupled effects of obliquity and eccentricity could result in a ``super habitable state'' with prospects for habitability even greater than Earth.

Finally, our assumption of synchronous rotation for planets with an obliquity is a simplification made in order to isolate the individual effects of obliquity and pCO$_2$ on climate.  \cite{Guerrero_2023} model the spin evolution of multi-planet systems, including the TRAPPIST-1 system. They show that high-obliquity planets are more likely to be sub-synchronous, i.e., with rotation periods longer than their orbital period. Specifically, the difference in the actual rotation period compared to the synchronous rotation period rapidly increases above 45$^\circ$ obliquity. \cite{Guerrero_2023} predict that the obliquity of TRAPPIST-1e is less than this threshold,
which implies the rotation period we assume is reasonable.
Similarly, \cite{Millholland_2024} uses N-body 
simulations to determine the obliquities of various planetary systems in resonant chains. They conversely predict that TRAPPIST-1e most likely has an obliquity of zero, mainly due to its low mutual orbital inclination. 

To better compare to upcoming observations, future work is needed to post-process the simulations \citep[e.g., with PSG,][]{Villanueva_2018} to determine if there are wavelength-dependent observable signatures of a non-zero obliquity. This includes both transmission and emission spectra to study the possible detectable impacts of obliquity with current and future observatories, including JWST, ELTs, MIRECLE, LIFE, and HWO.
\section{Conclusions} \label{Conclusion}
In this study, we explored the impacts of atmospheric composition and obliquity on the climate of TRAPPIST-1e, a temperate terrestrial planet. We present a grid of 3D climate models of TRAPPIST-1e with planetary obliquity ranging from 0$^{\circ}$ and 90$^{\circ}$ and pCO$_2$ ranging from 0.01 and 1 bar, assuming spin-synchronized rotation. 

We find that in general, pCO$_2$ has a dominant effect on the overall climate state relative to obliquity, but both pCO$_2$ and obliquity are important when considering the planet's dynamics and habitability. As pCO$_2$ increases, the surface temperature and cloud coverage generally become more globally uniform with reduced horizontal temperature contrasts, and the high-altitude mean wind speeds increase. As obliquity increases, the time-averaged equator-to-pole temperature contrast decreases and the mean surface temperature increases. 

As expected from \citep{Ohno_2019}, we find an abrupt shift in the atmospheric dynamics between 45$^{\circ}$ and 67.5$^{\circ}$ obliquity; given that above an obliquity of 54.5$^\circ$, the poles receive more irradiation than the equator on an annual average, affecting the mean climate. We find that the  transition in the instellation pattern with obliquity causes a dynamical shift wherein the super-rotating jet becomes sub-rotating at high obliquity, with regions of high cloud cover and surface temperature becoming located westward from the substellar point as opposed to eastward. 

Additionally, as the obliquity increases, the maximum time-mean cloud cover shifts poleward and westward.
Cloud cover directly affects the outgoing longwave radiation and thus the shape of the thermal phase curve by reducing the top-of-atmosphere outwelling flux in cloudy regions. In the simulated phase curves, the peak flux shifts from westward of the stellar point at low obliquity to eastward of the substellar point at high
obliquity. We find that the variation in phase curve offset with obliquity is largest for the cloudier cases with high pCO$_2$. We anticipate that obliquity could be inferred from observational phase curves if temperate rocky planets
are sufficiently warm
to have significant cloud coverage.
\vspace{5mm}

We thank Melissa Hayes-Gehrke and Eliza Kempton for their helpful feedback on an early draft of this work. The authors acknowledge the University of Maryland supercomputing resources (\url{https://hpcc.umd.edu}). This work was completed with resources provided by the University of Maryland Astronomy Department. We acknowledge generous support from the Heising-Simons Foundation as part of the 51 Pegasi b Fellowship Enhancements.
\bibliography{sources.bib}

\end{document}